# Keck Spectroscopy and HST Imaging of Field Galaxies at Moderate Redshift[1,2]

Duncan A. Forbes, Andrew C. Phillips, David C. Koo and Garth D. Illingworth

Lick Observatory, University of California, Santa Cruz, CA 95064

Received ___________ ;   accepted ___________

## ABSTRACT

We present 18 spectra, obtained with the Keck 10m telescope, of faint field galaxies ($19 < I < 22$, $0.2 < z < 0.84$) previously imaged by *HST's* WFPC2. Though small, our sample appears to be representative of field spirals with a magnitude–limit of $I \leq 22$. Combining the results from the spectral and imaging data, we have derived various quantitative parameters for the galaxies, including colors, inclinations, emission line equivalent widths, redshifts, luminosities, internal velocity information and physical scale lengths. In particular, disk scale lengths (with sizes ranging from ∼1–5 kpc) have been measured from fits to the surface brightness profiles. We have also measured internal velocities with a rest frame resolution of $\sigma = 55$ to 80 km s$^{-1}$ by fitting to the emission lines. The luminosity–disk size and luminosity–internal velocity (Tully–Fisher) relations for our moderate redshift galaxies are similar to the scaling relations seen for local galaxies, albeit with a modest brightening of ∼1 magnitude. The one bulge–dominated galaxy in our sample (at $z = 0.324$) has a relatively blue color, reveals weak emission lines and is ∼0.5 magnitude brighter in the rest frame than expected for a passive local elliptical. Our data suggest that galaxies at about half the age of the Universe have undergone mild luminosity evolution to the present epoch, but are otherwise quantitatively similar to galaxies seen locally.

*Subject headings:* galaxies: evolution – galaxies: formation – galaxies: structure – galaxies: fundamental parameters

## 1. Introduction

The nature and evolution of galaxies is an active research area in contemporary astronomy. Considerable telescope time has been spent pursuing this topic, yet it is still the subject of much debate due largely to the observational difficulties involved in collecting large, unbiased data sets of galaxies at faint magnitudes and to the lack of comparable local samples. The bulk of research has focused on number counts (to around B $\sim$ 26), redshift distributions (to B $\sim$ 24), clustering properties, and integrated colors (e.g. Koo & Kron 1992; Lilly et al. 1995). A few ground–based studies of the morphology of distant galaxies (to B $\sim$ 24) have been attempted (e.g. Lavery, Pierce & McClure 1992; Giraud 1992), but these are limited to spatial resolutions of $\geq 0.5''$. Clear advances in this area have been made by the *Hubble Space Telescope* using the first Wide Field and Planetary Camera (e.g. Dressler et al. 1994a; Griffiths et al. 1994a; Couch et al. 1994; Phillips et al. 1995a) and recently with the much improved WFPC2 (Dressler et al. 1994b; Forbes et al. 1994; Griffiths et al. 1994b; Cowie, Hu & Songaila 1995). Imaging with the *HST* allows morphological and structural information on subkiloparsec–size scales to be determined for distant galaxies.

An important tool for studying nearby galaxies via their internal kinematics is the Faber–Jackson relation for elliptical galaxies (Faber & Jackson 1976) and the Tully–Fisher relation for spiral galaxies (Tully & Fisher 1977). So far, there are only a handful of galaxies with redshifts greater than 0.1 for which internal velocity measurements have been published. Franx (1993a,b) derived internal velocity information for several galaxies in the cluster A665 (z = 0.18). At a slightly higher redshift (z $\sim$ 0.2), Vogt et al. (1993) measured a full rotation curve for field two spiral galaxies. Kinematic information on more distant galaxies would provide a powerful diagnostic of internal dynamics and allow galaxy evolution to be traced by mass as distinct from light.

Here we present a small data set from which we examine internal kinematics and sizes for galaxies at moderate redshift ($0.2 < z < 0.84$). Our data consist of 18 field galaxies with I band magnitudes between 19 and 22 ($\sim 20.5 <$ B $< 24.5$), for which we have high spatial resolution imaging from the WFPC2 camera of *HST* and moderate resolution spectra obtained on the Keck 10m telescope. These galaxies were originally selected from field galaxies in Medium Deep Survey (MDS) images (Griffiths et al. 1994a; Forbes et al. 1994). The images and spectra allow us to derive redshifts, velocity information, equivalent widths, Hubble classifications, inclinations, colors and physical scale lengths for these galaxies.

The median redshift of our sample ($\bar{z} = 0.48$) corresponds to a look–back time that is about half the age of the Universe and for which some scenarios predict significant evolution in galaxy populations (for a review, see Koo & Kron 1992). Of particular interest are the 'faint blue galaxies' that appear to dominate the number counts of field galaxies, with a factor of 3–5 above no–evolution predictions, by B $\sim$ 23 (e.g. Cowie, Songailia & Hu 1991). It has been suggested that these galaxies may have faded by several magnitudes (e.g. Cowie et al. 1991) or merged (e.g. Broadhurst, Ellis & Glazebrook 1992), so that they have 'disappeared' by the present epoch and thus do not have local counterparts. Our main result from this work, after comparing the size and internal kinematics with luminosity to those of local galaxies, is that *a sample of representative disk galaxies up to half the age of the Universe has undergone only a modest amount ($\sim 1^m$) of luminosity evolution.* We assume $H_o = 75$ km s$^{-1}$ Mpc$^{-1}$ and $q_o = 0$ throughout this paper.

## 2. Observations and Data Reduction

The galaxies described in this paper come from three high galactic latitude MDS WFPC2 fields, called MDS U1, U5 and U54. The first two fields were studied by Forbes et al. (1994), who classified over 200 field galaxies according to their morphological type and derived total I band magnitudes. The field MDS U54 consists of $3 \times 1800$s F606W (V) band and $3 \times 1200$s F814W (I) band images observed on 1994 July 4. These data were reduced, combined, classified and calibrated in a manner similar to that described in Forbes et al. (1994). A mosaic of the WFPC2 galaxy images is given in Fig. 1 (Plate **), and a summary of the measurements from the imaging data in Table 1.

The galaxies selected for follow–up spectroscopy were chosen so that two would lie on a single long–slit. They were not picked to have a particular morphology or angular size. For one galaxy in our sample (#3) the redshift was known from the work of Glazebrook et al. (1994), but its nature (super–starburst or gravitational lens) as discussed by Glazebrook et al. was uncertain. A discussion of whether our galaxies are typical of those expected in an I < 22 magnitude–limited sample is given in section 4.1. The spectroscopic data were obtained during 1994 September 8–10 using the W. M. Keck 10m telescope on Mauna Kea, Hawaii. We used the Low Resolution Imaging Spectrograph (LRIS; Oke et al. 1995) in long–slit mode with a $1.0''$ wide slit. The spectral dispersion was 1.28 Å/pix and the spatial scale $0.22''$/pix. The data were obtained under FWHM $\sim 0.8''$ seeing conditions and partial clouds. After locating the galaxies in the slit–guiding camera, the slit was placed so as to obtain spectra of at least two galaxies. A typical integration was a single 1800s exposure (see Table 2 for details), covering 4700Å to 7300Å . The FWHM instrumental resolution is R $\sim$ 1300, which corresponds to a rest frame instrumental velocity ($\sigma$ = FWHM / 2.35) of 55–80 km s$^{-1}$ for our sample galaxies.

The data were reduced using the IRAF package. The bias was removed and the data corrected for the amplifier gain between the two halves of the CCD. Several dome exposures, illuminated by a quartz lamp, were combined and fit using a low order polynomial. In the absence of suitable sky frames, this traces the response across the slit (low spatial frequency variations). After dividing by the polynomial fit, we created a 'dome flat' which was used to correct for pixel-to-pixel variations to the few percent level. The spectrum centroid and S distortion were traced out using the *apall* task. This task was also used to define the sky region near the spectrum. Geometric distortion along the wavelength direction was found to be very small ($< 0.1$ pixels) for most of our data. Tests on the most extreme cases showed that correcting for this small distortion had no effect on the redshift determination or width of the emission lines within our measurement errors (further discussion of errors is given below). After subtracting sky, from regions typically 20 pixels wide on either side of the nucleus, we wavelength–calibrated using the night sky lines. The data have not been flux calibrated. A total of 18 galaxy spectra have been extracted.

## 3. Results

### 3.1. HST Imaging

We have fitted elliptical isophotes to the galaxy images using the same method as Phillips et al. (1995a, 1995b). From this modeling, we measure the axial ratio of the outer isophotes, position angle on the sky, and a surface brightness profile. The major axis profile is then fit with an $r^{1/4}$ law and/or an exponential disk profile simultaneously to give scale lengths and surface brightnesses. Surface brightness profile fits for our sample are shown in Fig. 2. The galaxy morphology is classified according to the system of Forbes et al. (1994), i.e. we assign Hubble types where possible or 'b' for bulge–dominated, 'i' for intermediate and 'd' for disk–dominated systems (given in Table 1). This division correlates strongly with apparent magnitude, so that galaxies with $I < 21$ can generally be assigned familiar Hubble types, while those fainter are classified more crudely. In this paper we have amended the Forbes et al. (1994) morphological type of one galaxy (#7) from 'b' to 'i' as we feel this is a more accurate description of the visual morphology. For each galaxy, we have also noted any peculiar visual structure, such as a double nucleus or any asymmetries that may be due to a merger or interaction. On this basis, there appears to be several galaxies with physical associations (see Table 1 and section 4.4).

### 3.2. Keck Spectroscopy

For 17 of the 18 spectra, we have derived redshifts from both emission and absorption lines. In Fig. 3 we show each galaxy spectrum, smoothed to the instrumental resolution, and the location of identified lines. The measured redshifts for the galaxies lie within a range of z = 0.205 to z = 0.837, with random errors given by the rms variation from the individual lines (see Table 2). We also assign a 'quality class' which indicates the reliability of the quoted redshift, with quality class A (very secure redshift) given to 13 spectra, quality B (less secure, but deserves a high weighting) to 3, quality C (marginal redshift, should be given a low weighting) to 1 spectrum, and quality D (no redshift determined) for 1 spectrum.

The strong emission lines have been fit with a single Gaussian profile for H$\beta$ and [OIII] 5007Å, and a double Gaussian profile for [OII] 3726,3729Å doublet (with amplitudes allowed to vary, and a fixed rest–frame separation of 2.75Å). An example of a Gaussian fit to H$\beta$ and [OIII] for the Sb galaxy #12, is given in Fig. 4. We have also fit the non–blended sky lines at similar wavelengths to the emission lines with single Gaussians in order to derive the appropriate instrumental broadening. The observed emission line FWHM, after subtracting in quadrature the instrumental profile, is given in Table 2. For the unresolved lines, we give upper limits to the galaxy's internal velocity, of FWHM equal to 120 or 200 km s$^{-1}$. These limits were chosen after smoothing various high and low S/N emission lines by a prescribed amount, and then determining the level of broadening that was readily detectable. The main source of error in the FWHM determination is the quality of the Gaussian fit, which is typically $\sim 25\%$ based on differences in the individual emission line fits. For the equivalent width, it is the continuum level (error $\sim 35\%$). For 10 galaxies we were able measure the velocity broadening, whereas for the others we quote upper limits. The equivalent width of the [OII] line was measured for 13 galaxies. The measured FWHM line width (corrected for instrumental broadening) and the [OII] equivalent width are listed in Table 2 (they are the not corrected for inclination or redshift).

### 3.3. Derived Parameters

Combining results from the spectral and imaging data we have derived several physical parameters for our sample galaxies (see Table 3). In column 1 we give the Galaxy identification number. Columns 2 and 3 give the physical scale and absolute B magnitude ($M_B$) assuming $H_o = 75$ km s$^{-1}$ Mpc$^{-1}$ and $q_o = 0$. The absolute magnitude at zero redshift is calculated using the observed color with the tables of Frei & Gunn (1994), which give k–corrections as a function of redshift. The F814W filter is similar to the I band, while F606W lies between the V and R bands. The final rest frame magnitudes are estimated to be accurate to $\sim 0.2$ mag, with the dominant source of error being uncertainty in the k–correction. The magnitudes have not been corrected for either internal or galactic extinction (which is very small for these high–galactic latitude fields). The inclination, given in column 4, is calculated from the axial ratio of the outer isophotes, using $\cos^2 i = [(b/a)^2 - 0.04]/0.96$ (Rubin, Whitmore & Ford 1988). We estimate an inclination error in degrees of 10% for $\sin i \geq 60°$ and 15% for $\sin i < 60°$. Galaxy sizes and brightnesses (columns 5–9) are derived from the F814W (I) band images, so as to minimize the effects of obscuring dust and young star formation. The half light radius measured following the method of Phillips et al. (1995b), gives a *model–independent* measure of the galaxy light profile. Fitting the I band surface brightness profile with $r^{1/4}$ and/or exponential disk profile allows us to derive bulge and disk model parameters. We have adopted errors of $0.15^m$ in the surface brightnesses and 15% in the length scales (20% for the effective radii, $r_{eff}$, as it is somewhat less reliable) based on the simulations of Phillips et al. (1995a). Column 10 gives the bulge-to-disk ratio from the profile fits. Column 11 gives the emission line velocity width in terms of sigma ($\sigma$ = FWHM / 2.35) of the Gaussian. These final velocities have been corrected for instrumental broadening, (1 + z) redshift effect, sin $i$ inclination and spatial extent of the spectrum (see section 4.6). The quoted error includes an estimate from these different sources, but is dominated by the error in fitting a Gaussian profile to the emission line. For the single elliptical galaxy in our sample, we only made corrections for instrumental and redshift effects. The [OII] equivalent width, in column 12, is corrected for redshift. Here the error is dominated by the uncertainty in determining the continuum level of the emission line.

The 17 galaxies with redshifts range from z = 0.205 to 0.837. The absolute B magnitudes listed in Table 3 range from $M_B$ = –18.4 to –21.3. For comparison, an $L^*$ galaxy has $M_B$ = –20.1. Thus our sample ranges from about 1.5 magnitudes fainter than $L^*$ to about 1.5 magnitudes brighter than $L^*$.

## 4. Discussion

### 4.1. Is our Sample Representative of Faint Field Galaxies ?

As mentioned above, the galaxies for which we have obtained spectra come from a larger I < 22 magnitude–limited sample, which is being studied for its structural and photometric properties (e.g. Forbes et al. 1994). Selection effects are crucial if one is studying galaxy distributions (e.g. number counts), but less so when comparing intrinsic properties of individual galaxies, as done in this paper. Nevertheless we still wish to know whether our sample is representative of faint field galaxies.

The median redshift of our sample is $\bar{z} \sim 0.48$. The ground–based I < 22 redshift surveys of Lilly et al. (1995) and Tresse et al. (1993) find a median redshift of $\bar{z} \sim 0.6$, i.e. roughly comparable to that of our sample. In figures 5 and 6, we compare the color and half light radius to the somewhat larger sample of Phillips et al. (1995b). Phillips et al. confirmed that their WFPC2 selected sample was essentially 100% complete, over a small sky area, to I < 21.7. Of their 64 galaxies, 3 are in common with our sample. In comparing the total V–I color vs total I magnitude, we find that our sample has a similar distribution in V–I color to that of Phillips et al. between I = 19 and 21, but at fainter magnitudes we may be systematically biased against the reddest galaxies (see Fig. 5). These galaxies tend to be the bulge–dominated systems. Next, we compare the half light radii from Phillips et al. with our sample in Fig. 6, and find little difference.

Finally, we examine the morphological mix of our sample. Morphological types from WFPC2 are available for two hundred galaxies in Forbes et al. (1994). This sample is essentially complete to I < 22. Of the 203 galaxies, 10 are in common with our current sample. Figure 7 shows that the late type galaxies are fairly well represented in our current sample, but we seem to be deficient in early type/bulge–dominated galaxies relative to the large magnitude–limited sample (as suggested by the V–I color distribution). As we separate bulge-dominated galaxies from the analysis that follows, this difference in morphological mix shouldn't affect our conclusions. Although our sample is small, and in no way complete, it appears to be a reasonably fair and representative sample of field disk galaxies with I < 22.

## 4.2. The Elliptical Galaxy

Our sample contains one example of a bulge–dominated galaxy (#17). Although we have classified it as an elliptical, we can not rule out the presence of a weak disk (i.e. it could be an S0 galaxy) Visually, the galaxy reveals asymmetric isophotes. The inner parts of the surface brightness profile are well fit by a pure $r^{1/4}$ law (see Fig. 2). We derive an $r^{1/4}$ effective radius that is the same as the half light radius (2.6 kpc). This value can be compared to nearby ellipticals using the size–luminosity relation found by Binggeli et al. (1984). On this basis we would predict an effective radius, if it followed the local scaling relation, of 2.2 kpc (using $H_o = 75$). This agrees with the measured effective radius of 2.6 ± 0.5 kpc, and this is without taking into account the intrinsic scatter in the local relation. Similarly, if we use the effective radius to predict the internal velocity $\sigma$ using a projection of the fundamental plane as described by Guzmán et al. (1993), then we predict $\log \sigma = 2.26$. We measure an emission line velocity width of $\log \sigma = 1.91 \pm 0.48$, which is again consistent, within the measurement errors, to that of a local elliptical with ($r_{eff} = 2.6$ kpc). The absorption lines appear to have a similar velocity width.

The spectrum is interesting since, besides having the expected lines for an elliptical (i.e. Ca II H+K, the G band and Fe I 4384), it also reveals Balmer absorption lines, and [OII] 3727 and H$\beta$ in emission. These emission lines may suggest that the galaxy has undergone a recent episode of star formation, and may be a rare example of an active field elliptical (poststarburst ellipticals are usually found in clusters). We measure a total color V–I = 0.92, which is about $0.5^m$ bluer than that predicted (V–I $\sim$ 1.45) for a passive local elliptical at redshift z = 0.324 (Frei & Gunn 1994). The blue color could be due to recent star formation. Thus although this galaxy has structural parameters (i.e. $r_{eff}$ and $\sigma$) similar to those expected for a local elliptical, we find evidence for some additional brightening.

### 4.3. [OII] Equivalent Widths

For moderate redshift galaxies, the [OII] 3727 line has been used by several workers as an indication of high star formation (SF) rates (e.g. Couch & Sharples 1987; Broadhurst, Ellis & Shanks 1988). Although less accurate for quantitative studies, [OII] provides a reasonable substitute for H$\alpha$ since stellar H$\beta$ line absorption may affect H$\beta$ measurements and [OIII] 5007 is more sensitive to excitation variations (Kennicutt 1992). Other star formation diagnostics, such as Balmer absorption lines or the 4000 Å break, sample star formation over the past several Gyrs.

The rest-frame [OII] equivalent widths for our sample are given in Table 3. If we divide the sample by *intrinsic* color into two bins at B–V = 0.55, we find that the mean equivalent width for the 'red' galaxies is 19Å whereas the 'blue' galaxies have a mean value of 59Å . As expected, intrinsically blue galaxies have higher star formation rates. Equivalent widths greater than 20Å suggest a SF rate of a few solar masses per year for an L$^*$ galaxy (Kennicutt 1992). We have 7 such galaxies, some of which have an asymmetric visual morphology, but so do many of the galaxies with [OII] equivalent widths less than 20Å .

### 4.4. Physical Associations

There are several galaxies in our sample with similar redshifts. They are numbers 6 and 7 (at z $\sim$ 0.595), and 4, 5 and 1 (at z $\sim$ 0.477). The former galaxies consist of a barred spiral, with a possible foreground star superposed on it, paired with a faint galaxy classified as intermediate type (see Fig. 1). Given the small projected separation of 27 kpc and the distorted outer isophotes of the spiral, it appears that the two galaxies are tidally interacting. The latter three galaxies represent a double nucleus galaxy (presumably in the late stage of a merger), and two $\sim$L$^*$, undisturbed spiral galaxies. The projected separation from #5 to #4 is 110 kpc and 460 kpc to #1.

We note that some apparent associations in our sample, are merely projections on the sky, i.e. they appear close, with similar angular sizes and yet are well separated in redshift space. Such an example is shown in Fig. 1 by galaxies #15 (z = 0.566), 16 (z ∼ 0.4 ?; see the appendix), 17 (z = 0.324) and 18 (z = 0.754). Some of these galaxies show evidence for asymmetries or distortions, which might be thought due to an interaction with one of the other galaxies that is projected nearby. As they are at different redshifts this is not the case, but it does serve to highlight the risks of classifying interacting galaxies on the basis of visual appearance alone. Estimates of the frequency of interacting and merging galaxies from imaging (without knowing the redshifts of both interacting systems) are likely to be overestimated. To illustrate this, if we take the extreme view that all galaxies with asymmetries and multiple nuclei are due to a merger or interaction (as has been done at times) then our small sample would suggest 11/17 galaxies could be so classified. With redshifts and imaging data, we would reduce this to only 3–4 galaxies which we would consider to be clearly merging or interacting with another galaxy in the sample. This comparison doesn't include companion galaxies that are not in our redshift sample, or evolved systems that show asymmetries but no other evidence of the secondary galaxy.

## 4.5. Luminosity – Size Relation

Since Freeman (1970) observed that spiral galaxies have a constant disk central surface brightness of $\mu_B = 21.65 \pm 0.3$, there has been much debate concerning the interpretation and reliability of this result. The most comprehensive study to date is that of de Jong (1995), who shows that there is no single preferred value but a range of $\mu_B$ that varies with Hubble type. He finds, for spirals (RC2 types 1–6) with semi–transparent disks, that $\mu_B = 21.45 \pm 0.76$ mag arcsec$^{-2}$. We have adopted this value for the local central surface brightness, which can be expressed in terms of a local luminosity–size scaling relation.

Assuming the total light from a pure exponential disk is $L_T = 2\pi\theta_d^2 L_o$, where $\theta_d$ is the angular disk scale length and $L_o$ is the central surface brightness (in $L_\odot$ arcsec$^{-2}$), we can derive the local luminosity–disk size scaling relation to be:

$$\log(r_d/\text{kpc}) = -0.2 \; M_{B,disk} - 3.42 \; (\pm \; 0.15)$$

We have calculated the disk magnitude ($M_{B,disk}$), for our sample, by subtracting the light in the bulge ($< 0.2^m$ in all cases) from the total magnitude in Table 3. We have not made any correction for inclination or internal absorption (which tend to largely cancel each other in calculating face–on surface brightnesses).

In Fig. 8 we show the physical disk scale length against rest frame absolute B magnitude of the disk for our sample galaxies and the local scaling relation for spirals. Most of our sample galaxies fall slightly below the mean relation, i.e. to smaller scale lengths and/or to brighter magnitudes. A linear regression fit, inversely weighted by the error in disk scale length, gives:

$\log(r_d/\text{kpc}) = -0.15 \pm 0.05 \; M_{B,disk} - 2.58 \pm 1.0$

Thus our sample is offset to lower scale lengths with a flatter slope, but with formal errors that are consistent with the local relation. If we fix the slope to a value of –0.2, then the change in the other coefficient relates to a change in central surface brightness only. This gives a brightening of $0.85^m$, corresponding to $\mu_B = 20.6 \pm 0.2$ mag arcsec$^{-2}$ for the sample as a whole. We note that surface brightnesses are unaffected by changes in $q_o$. If we restrict our sample to just those that are clearly spiral in nature (filled symbols in Fig. 8), then both the slope and intercept are closer to the local relation values. Again for a fixed slope, the spirals are brightened by $\Delta\mu_B \sim 0.6^m$ giving $\mu_B = 20.9 \pm 0.1$. Thus our sample suggests that deviations from the local scaling relation are strongest in the very late type/irregular galaxies (although in some cases this is probably because the disk scale length is poorly defined). Schade et al. (1995) have recently examined the luminosity–size relation for 32 galaxies with redshifts $0.5 < z < 1.2$. Scale lengths have been measured from WFPC2 images and redshifts obtained from CFHT spectra. They find a value of $\mu_B = 20.3 \pm 0.2$ for galaxies with obvious spiral structure, and find no statistical difference between these spirals and the small featureless objects or irregulars. However, they also note the difficulty of quantifying the meaning of a disk scale length for the small and irregular galaxies. Their data for more distant galaxies ($\bar{z} \sim 0.75$) suggests a brightening of $\Delta \mu_B \sim 1.2^m$, i.e. slightly higher than our estimate for $\bar{z} \sim 0.5$ galaxies. When we divide our sample into intrinsically blue and red galaxies at B–V = 0.55 (circles and triangles respectively in Fig. 8), we find $\mu_B = 20.8 \pm 0.3$ for the blue galaxies and $\mu_B = 20.4 \pm 0.3$ for the red galaxies. Although the errors are large, there is an indication that the blue galaxies deviate less from the local relation than do the red galaxies. It is possible that star formation has conspired to move points parallel to the local relation, i.e. low luminosity systems may have star formation preferentially in their outer parts, so that the brightening is accompanied by an increased scale length. If this were the case, we might expect the higher luminosity galaxies in our sample to be dominated by blue galaxies – which is not seen.

Additional support for the modest brightening of distant field galaxies comes from the ground-based study of Colless et al. (1994). Their sample is at a slightly lower median redshift ($\bar{z} \sim 0.3$) than ours but contains a comparable number of galaxies. Their seeing was sub-arcsec ($0.5''$–$1.0''$), but it was not sufficient for a direct visual classification of the galaxy's morphological type. Furthermore, they could not decompose the surface brightness profiles into bulge and disk combinations, as we have done. With the limited number of resolution elements, they were only able to fit either a disk *or* a bulge profile. This will introduce a small bias towards smaller scale lengths for early type spirals. They measured a total of 19 galaxy scale lengths and 7 upper limits. After excluding the two

bulge–dominated systems, we also show their data in Fig. 8 (after correcting to $H_o = 75$ km s$^{-1}$ Mpc$^{-1}$ and $q_o = 0$). Their data also follow the same general trend as ours, slightly below the local relation and with a flatter slope. A similar conclusion (that disk galaxy sizes have evolved little) was reached by studying somewhat brighter, less distant ($\bar{z} \sim 0.2$) field galaxies from WFPC1 imaging (Mutz et al. 1994; Phillips et al. 1995a).

We conclude that *spiral galaxies at moderate redshift follow a similar luminosity–size relation to local galaxies, albeit with a modest brightening of ∼1 magnitude.*

### 4.6. Luminosity – Velocity Relation

Next we examine the relationship between a galaxy's internal velocity and its total luminosity, and compare it to those of local galaxies. There does not yet exist a systematic and complete sample of local field galaxies studied by their internal velocity properties. Most studies of the Tully–Fisher relation have concentrated on well–behaved, non–interacting, non–barred galaxies so as to reduce the scatter about the relation for distance–determination purposes. In Fig. 9 we show the local scaling relations of Rubin et al. (1985) for normal (i.e. non–interacting, non–barred) field spirals. Their $V_{max}$ of the rotation curve has been converted into sigma of a Gaussian assuming that the FWHM of the Gaussian represents the full range in internal velocity, i.e. $\sigma = 2 \times V_{max}/2.35$. This is roughly analogous to integrated HI velocity widths, $W_{50}$. The Rubin et al. relations are derived from H$\alpha$ rotation curves and have a dispersion of $\sim 0.5^m$ per Hubble type. The HI derived Tully–Fisher relation (e.g. Pierce & Tully 1988) falls between the Rubin et al. Sa and Sc galaxy relations. We would expect Sd/Irr galaxies, with a lower velocity for a given luminosity, to lie slightly below the Rubin et al. relation for Sc spirals. The measured velocity width in barred galaxies will depend on the relative orientation of the bar and slit. We also show in Fig. 9, the local relation for HII galaxies from Telles (1995). Such galaxies are dominated by a global starburst and have M$_B$ up to –21.0 (Telles & Terlevich 1993). The internal velocities for the HII galaxies are derived from H$\alpha$ emission line widths. Thus at a given luminosity, there will be a large range of internal velocities for a sample of randomly selected local field galaxies, unlike disk size which reveals a relatively tight relationship with luminosity.

By using local relations that are determined from ionized gas, we will avoid any systematic differences from those derived using neutral hydrogen gas. However we must assume that the spatial distribution of [OII], H$\beta$ and [OIII] derived velocities are similar to that of H$\alpha$. In Fig. 9, we express internal velocity in terms of $\sigma$ of the Gaussian (= FWHM / 2.35) from the emission line widths. Gaussians are a good approximation to the emission line profiles. After measuring the emission line velocity widths, we have made the usual corrections for instrumental broadening, redshift effect, and inclination.

As our galaxies tend to be somewhat larger than the slit width, the orientation of the slit is important. Thus we need to make an additional correction for the limited spatial coverage of the slit along the galaxy's major axis. The measured velocity needs to be increased if the spatial extent of the spectrum does not probe to sufficiently large galactocentric radii in order to reach each galaxy's maximum rotation velocity (V$_{max}$). Fortunately, galaxy rotation curves are relatively flat beyond the central regions, and so any corrections are small. This effect can be crudely quantified using the parameterization of local galaxy rotation curves found by Persic & Salucci (1991), based on the fact that outer rotation curve gradients vary systematically with galaxy luminosity. They found the radial dependence of rotation velocity to be:

$$\frac{V(r)}{V_{max}} = 1 + (0.12 - 0.24 log \frac{L_B}{L_{B*}})(\frac{r}{2.2 r_d} - 1)$$

We recognize that using this relation assumes that distant galaxies have similar rotation curve forms to local galaxies, but as this correction is typically ∼8% for our sample, it does not affect the general conclusions. The larger corrections are applied to those few galaxies with slits close to the minor axis. Such galaxies in our sample tend have velocity upper limits. The final corrected velocities are listed in Table 3.

For two of our galaxies, the emission lines of H$\beta$ and [OII] are extended by about 10 pixels from the spectrum centroid thus allowing us to derive spatial velocity information, or rotation curves, and hence measure $V_{max}$ directly for comparison with the velocity derived from emission line widths. The two galaxies (# 4 and 5) are a double nucleus system and a regular–looking Sb spiral, both at redshift z ∼ 0.48. Applying the same redshift and inclination corrections, and assuming $\sigma = 2 \times V_{max} / 2.35$, the rotation curves give a sigma value of log $\sigma$ = 2.11 and log $\sigma$ = 2.23 for galaxies 4 and 5 respectively. In Table 3 we give values of 2.10 ± 25% and 2.08 ± 30% respectively. Thus, the emission line velocity widths are similar, but perhaps slightly lower, than those derived from the emission line rotation curves. A full description of the rotation curve derivation for these galaxies, and others from a different study, will be presented by Vogt, Forbes & Phillips (1995).

We find that our sample galaxies fill the region between local Sa and HII galaxies. Most of the spiral galaxies are barely consistent with the Sc galaxy relation. If we take the six spirals with detections, and the same slope as the Rubin et al. relation, we find that the brightening is $\Delta$ $M_B$ ∼ $0.4^m$ with respect to the Sc relation. Of the galaxies that lie close to the HII galaxy relation, most of these have intrinsic blue colors, often with very high [OII] equivalent widths (i.e. they are likely to be starbursts). One exception is the barred spiral (#6) for which the slit was placed directly along the bar, which may have led to a systematically lower velocity width (van Albada & Roberts 1981). The starbursting galaxies may be similar to the moderate redshift compact narrow emission line galaxies (CNELGs) described by Koo et al. (1995a), which they showed are equivalent to luminous examples of local HII galaxies. The presence of a starburst could lead to a systematically lower emission line width if a small number of low velocity star forming complexes (giant HII regions) dominate the ionized gas in the central regions. In this case, the line width could have a significant contribution from the internal motion induced by stellar energy release within the HII regions. Some support for this comes from Telles & Terlevich (1993) in which they find HII galaxies may have emission line widths that are up to a factor of 2 (0.3 in the log) smaller than those derived from HI 21cm measurements. Other possibilities also remain, such as a small spatial extent for the optical data compared to the HI data. These effects would move points diagonally to the lower right in Fig. 9.

We also plot data for two disk systems from Vogt et al. (1993), after converting their R band magnitudes (Vogt 1995) to $M_B$ using the Frei & Gunn (1994) tables and assuming both to be late–type spirals (they did not have the benefit of HST imaging). The first galaxy (z = 0.211) is described as 'elongated with a small bulge', and has very strong emission lines. It lies just above the HII relation in Fig. 9. The second (z = 0.201) is an 'elongated' cluster galaxy, presumably also a spiral, that lies close to the relation for Sc galaxies. These two galaxies lie within the locus of our sample, and are also consistent with some brightening relative to local galaxies.

In summary, we find that our moderate redshift field galaxies fall within the region defined by local spirals and HII galaxies in terms of their internal velocities. Many of the spiral galaxies in our sample lie close to the local relation for Sc galaxies, with only modest brightening. However, a sizable fraction of the sample lie closer to the local relation for HII galaxies. These galaxies tend to be starbursting galaxies as indicated by their blue colors and high [OII] equivalent widths. For these galaxies, it is important to check the emission line velocity widths against those determined directly from rotation curves. So although our

luminosity–velocity relation doesn't provide us with such a clear–cut interpretation as the luminosity–size relation, it does suggest that some fraction of moderate redshift galaxies have velocity widths characteristic of local galaxies with $\sim 1^m$ of brightening.

### 4.7. The Nature of the Faint Blue Galaxies

One of the major outstanding puzzles of cosmology is the nature of the blue galaxies that appear to dominate the number counts, by a factor of 3–5 above non–evolution predictions, by B $\sim$ 23. For example, Cowie et al. (1991) presented redshifts for 22 galaxies with B $\leq$ 24 and claimed that counts at these magnitudes were "...dominated by a population of small blue galaxies..." that is not seen locally. The current explanations for these blue galaxies include 1) an entirely new population at moderate redshifts that has disappeared by today (Cowie et al. 1991; Babul & Rees 1992), 2) mergers and associated star formation that have distorted the galaxy luminosity function compared to that at the present epoch (Broadhurst et al. 1992), or 3) that uncertainties in the local galaxy luminosity function are not known well enough to require alternative explanations yet (Koo & Kron 1992). The first two of these scenarios would predict that there exists a *large* population of galaxies at moderate redshift that have 'disappeared' by the present epoch and thus do not have local counterparts.

Recently, Cowie et al. (1995) have presented deep WFPC2 F814W images of faint field galaxies. The main result of their paper is the identification of a completely new morphological class of galaxy called 'chain galaxies'. These galaxies are fainter than our magnitude limit of I = 22, and are very blue with a linear, beaded morphology. Cowie et al. suggests that they have redshifts z $\sim$ 1.5. If spectra are obtained for these galaxies, it will be useful to compare their properties with the luminosity–size and luminosity–internal velocity relations to assess their true nature.

As discussed in section 4.2, our sample, although small, appears to be representative of disk galaxies at B $\sim$ 23, i.e. the regime of the faint blue galaxy excess. If the excess is a factor of 3–5 above non–evolutionary predictions (Cowie et al. 1991) then we would expect about 75% or 13–14 galaxies in our sample to belong to such an excess population. What then can we say about such galaxies from our sample ? By examining their disk sizes and internal velocities, we find that many of our sample galaxies show evidence for modest ($\Delta$ M$_B$ $\sim 1^m$) evolutionary brightening relative to the present epoch, but otherwise have quantitative parameters that are similar to the local scaling relations. The single bulge–dominated galaxy in our sample appears to be structurally similar to local ellipticals but is $\sim 0.5^m$ bluer than expected for a passive local elliptical and reveals weak emission lines. Our general findings are supported by the WFPC2 study of Phillips et al. (1995b), which is complete to I $<$ 21.7 (B $<$ 24), and finds that the angular size vs apparent magnitude relationship is consistent with mild luminosity evolution of the local galaxy populations. As our sample is small, it is premature to draw strong conclusions regarding the nature of faint galaxy evolution. Nevertheless, we do not see evidence for a dominant population of galaxies at moderate redshift that have no local counterparts, once a modest amount of fading has occurred. This is consistent with the recent results from deep surveys of Lilly et al. (1995).

## 5. Concluding Remarks

In the near future, we expect much new data to address the issue of galaxy sizes and internal kinematics at moderate redshift, using the more efficient multi–slit mode of the LRIS spectrograph on the Keck telescope (e.g. Koo et al. 1995b) and with the AUTOFIB multi–fiber instrument on the AAT (Rix, Guhathakurta & Colless 1995). The combination of full rotation curves and surface brightness profiles will allow us to examine the radial

– 22 –

dependence of mass and M/L in these distant galaxies (e.g. Persic & Salucci 1990; Forbes 1992). With larger samples, and thus better statistics, additional aspects of faint galaxy evolution can be addressed, such as the morphology–density relation at moderate redshift, clustering properties of field galaxies and the change in galaxy properties with redshift.

In summary, for a small but representative sample of moderate redshift ($0.2 < z < 0.84$) field galaxies, we present spectra from the Keck telescope and imaging from *HST's* WFPC2. From these data, we have derived various quantitative physical parameters, including scale lengths from surface brightness profiles and internal velocities with a resolution of $\sigma = $ 55–80 km s$^{-1}$. We examine the relationship between galaxy luminosity and both disk scale length and internal velocity. Together these data suggest that many spiral and disk galaxies with redshifts $\sim 0.5$ have undergone modest $\Delta$ M$_B$ $\sim 1^m$ luminosity evolution. Otherwise they have values characteristic of local galaxies. The single elliptical galaxy in our sample has structural parameters (i.e. r$_{eff}$ and $\sigma$) consistent with those for a local elliptical but is much bluer in the rest frame than expected for a passive elliptical. Although we have found evidence for mild luminosity evolution at moderate redshifts, we *do not* see evidence for dominant population of galaxies that has completely disappeared by the present epoch, as suggested by some galaxy evolutionary scenarios.

We thank S. Faber, R. Guzmán, T. Bida and the Keck observatory staff for help with the observations, and N. P. Vogt, G. Wirth and P. Guhathakurta for useful discussions. We also thank the referee for many helpful comments. This research was funded by grants GO-2684.04-87A, GO-2684.05-87A, AST91-20005 and AST88-58203.

**References**

Babul, A., & Rees, M. J. 1992, MNRAS, 255, 346

Broadhurst, T., Ellis, R. S., & Shanks, T. 1988, MNRAS, 235, 827

Broadhurst, T., Ellis, R. S., & Glazebrook, K. 1992, Nature, 355, 55

Binggeli, B., Sandage, A., & Tarenghi, M. 1984, ApJ, 304, 305

Colless, M. M., Schade, D., Broadhurst, T. J., & Ellis, R. S. 1994, MNRAS, 267, 1108

Couch, W. J., & Sharples, R. M. 1987, MNRAS, 229, 423

Couch, W. J., Ellis, R. S., Sharples, R. M., & Smail, I. 1994, ApJ, 430, 121

Cowie, L. L., Songailia, A., & Hu, E. M. 1991, Nature, 354, 400

Cowie, L. L., Hu, E. M. & Songailia, A. 1995, AJ, 110, 1576

de Jong, R. S. 1995, PhD Thesis, Groningen

Dressler, A., Oemler, A., Butcher, H. R., & Gunn, J. E. 1994a, ApJ, 430, 107

Dressler, A., Oemler, A., Sparks, W. B., & Lucas, R. A. 1994b, ApJ, 435, L23

Faber, S. M., & Jackson, R. E. 1976, ApJ, 204, 668

Forbes, D. A. 1992, A & AS, 92, 583

Forbes, D. A., Elson, R. A. W., Phillips, A. C., Koo, D. C., Illlingworth, G. D. 1994, ApJ, 437, L17

Franx, M. 1993a, ApJ, 407, L5

Franx, M. 1993b, PASP, 105, 1058

Freeman, K. 1970, ApJ, 160, 811

Frei, Z., & Gunn, J. E. 1994, AJ, 108, 1476

Giraud, E. 1992, A & A, 257, 501

Glazebrook, K., Lehar, J., Ellis, R., Aragon–Salamanca, A., & Griffiths, R. 1994, MNRAS, 270, L63

Griffiths, R. E., et al. 1994a, ApJ, 437, 67

Griffiths, R. E., et al. 1994b, ApJ, 435, L19

Gronwall, C., & Koo, D. C. 1995, ApJ, in press

Guzmán, R. Lucey, J. R., & Bower, R. G. 1993, MNRAS, 265, 731

Kennicutt, R. C. 1992, ApJ, 388, 310

Koo, D. C., & Kron, R. G. 1992, ARAA, 30, 613

Koo, D. C., Guzmán, R., Faber, S. M., Illingworth, G. D., & Bershady, M. A., Kron. R. G., & Takamiya, M. 1995a, ApJ, in press

Koo, D. C., et al. 1995b, in preparation

Lavery, R. J., Pierce, M. J., & McClure, R. D. 1992, AJ, 104, 2067

Lilly, S. J. et al. 1995, preprint

Mutz, S., et al. 1994, ApJ, 434, L55

Oke, J. B., et al. 1995, PASP, 107, 375

Persic, M., & Salucci, P. 1990, ApJ, 247, 349

Persic, M., & Salucci, P. 1991, ApJ, 368, 60

Phillips, A. C., Bershady, M. A., Forbes, D. A., Koo, D. C., Illingworth, G. D., Reitzel, D. B., Griffiths, R. E., & Windhorst, R. A. 1995a, ApJ, 444, 21

Phillips, A. C., et al. 1995b, in preparation

Pierce, M. J., & Tully, R. B. 1988, ApJ, 330, 579

Rix, H. W., Guhathakurta, P., & Colless, M. M. 1995, in preparation

Rubin, V. C., Burstein, D., Ford, W. K., & Thonnard, N. 1985, ApJ, 289, 81

Rubin, V. C., Whitmore, B. C., & Ford, W. K. 1988, ApJ, 333, 522

Schade, D., Lilly, S. J., Crampton, D., Hammer, F., Le Fevre, & O., Tresse, L., 1995, preprint

Telles, E. 1995, personal communication

Telles, E., & Terlevich, R. 1993, Astro. & Sp. Sc., 205, 49

Tresse, L., Hammer, F., Le Fevre, O., & Proust, D. 1993, A & A, 277, 53

Tully, R. B., & Fisher, R. 1977, A & A, 54, 661

van Albada, G. D., & Roberts, W. W. 1981, ApJ, 246, 740

Vogt, N. P., Herter, T., Haynes, M. P., & Courteau, S. 1993, ApJ, 415, L95

– 25 –Vogt, N. P. 1995, personal communication

Vogt, N. P., Forbes, D. A., & Phillips, A. C. 1995, in preparation

## Appendix: Individual Galaxies

**030505.0–001143** (#1) This is a good example of a near face–on spiral at redshift 0.477. It obeys both the luminosity–disk size and luminosity–velocity width scalings for local galaxies.

**030504.9–001138** (#2) Although projected on the sky close to galaxy 030505.0–001143, it is at a higher redshift. This galaxy is a late–type galaxy with an extremely high rest–frame [OII] equivalent width (150Å ), i.e. currently undergoing a burst of star formation. It is the faintest galaxy in our sample with I = 21.9. It interesting to note that this galaxy has fairly similar half light radius to galaxy #1, and yet one would expect that they have quite different values from examining Fig. 1 by eye. This is largely due to the 2 mag difference in surface brightness of the two objects which is not obvious in the image.

**030501.3–001039** (#3) We chose this galaxy based on its peculiar morphology and its unknown nature (Glazebrook et al. 1994 suggested either a super–starburst galaxy or a gravitational lens). The redshift was known from the work of Glazebrook et al. (the only one in our sample). The spectrum, from the main part of the lower ring, supports the starburst origin with strong, unresolved lines typical of an HII region and a large [OII] equivalent width. Examination of the 2-D spectrum suggests that there is small velocity shift across the system, making the gravitational lens interpretation less likely. The disk scale length and surface brightness are less reliable for this peculiar galaxy than for others.

**030458.0–001135** (#4) This galaxy shows a clear double nucleus, separated by ∼1.0$''$ (5.3 kpc), and is presumably in the late stage of a merger. The measured line width reflects the two velocity components. An exponential fit was made to the isophotes giving $r_{disk}$ = 1.8 kpc, but the scale size is not very meaningful in this case.

**030459.2–001146** (#5) At the same redshift as the double nucleus galaxy (z = 0.477). It has a high luminosity ($M_B$ = −21.3) and has the largest disk scale length ($r_{disk}$ = 5.1 kpc) in our sample.

**030503.4–001010** (#6) We classified this galaxy as a barred spiral. The photometry is somewhat uncertain, as we have attempted to exclude the possible foreground star which lies just off the nucleus. The slit is parallel to the galaxy bar, so in this case we may have a significant contribution from non–circular motions.

**030503.3–001015** (#7) This galaxy appears to be a physical companion of the barred spiral (030503.4–001010) since both have a redshift of ∼0.595 and a separation of 27 kpc. The morphological type is somewhat uncertain; we have classified it as an intermediate type but the surface brightness profile is well fit by a single exponential disk.

**010958.5–022724** (#8) We have classified this galaxy as an asymmetric Sb, as its appearance is clearly distorted from a typical Sb spiral. It has a relatively high velocity width and [OII] equivalent width.

**010958.1–022740** (#9) This is the closest galaxy in our sample (with a redshift of 0.205) and the least luminous. It appears to be undergoing a merger or interaction with a small galaxy, and so we have classified it as a double nucleus system.

**010957.4–022807** (#10) This galaxy has a low surface brightness, and appears to show a spiral arm structure. It is very blue with strong [OII] emission, suggesting recent star formation.

**171220.8+333559** (#11) At a redshift of 0.837, this is the most distant galaxy in our sample. It is also extreme in its rest–frame [OII] equivalent width (∼150Å ). Visually it appears to be asymmetric.

**171221.4+333556** (#12) This is a nearby (z = 0.256), asymmetric and blue Sb spiral. Its luminosity is close to L*. The [OII] 3727 line is blueward of our wavelength coverage, but the visual morphology and blue color suggest a recent burst of star formation across the disk.

**171227.1+333549** (#13) Images of this galaxy show distortions with a possible faint companion. The galaxy is distant (z = 0.759) and luminous ($M_B$ = –21.1).

**171227.0+333558** (#14) The morphological type of this galaxy is somewhat uncertain; we have classified it as an intermediate type but the surface brightness profile is well fit by a single exponential disk. The galaxy is also quite blue.

**171229.5+333626** (#15) In this case, the slit lies close to the minor axis of the galaxy and a fairly large correction to the measured velocity width was required. The image suggests spiral arms and ongoing star formation. The luminosity ($M_B$ = –20.3) and scale length ($r_{disk}$ = 4.1 kpc) are close to to L* and the average disk scale length for local galaxies of 3.5 kpc quoted by Mutz et al. (1994).

**171229.5+333634** (#16) We have not managed to determine a convincing redshift for this galaxy (the only one in our sample without a redshift). Comparing the angular half light radius and apparent I magnitude with the latest mild luminosity evolution models of Gronwall & Koo (1995) suggests that z ∼ 0.4. This is consistent with a 4000Å break in the spectrum at ∼5600Å . The resulting absolute magnitude $M_B$ = –18.9 would place it at the low–luminosity end of our sample. The galaxy shows evidence for a weak bar along the major axis, and the outer isophotes are clearly distorted, suggesting an interaction, possibly with a faint companion just visible in Fig. 1.

**171229.3+333636** (#17) This is the only clear case of a bulge–dominated galaxy in our sample (z = 0.324), and has been discussed in detail in section 4.2.

**171229.9+333644** (#18) Close to face–on, this galaxy shows asymmetric structure and possible off–nucleus star formation. The rest frame [OII] equivalent width is 34Å . It is a highly luminous ($M_B$ = –21.3), distant (z = 0.754) galaxy.

– 29 –

**Figure Captions**

**Figure 1** (Plate **) Grey scale mosaic of the WFPC2 images of our galaxy sample. The display is logarithmic. Each subimage box has the same scale, with the small boxes being 100 pixels ($10''$) on a side. The galaxies are numbered as in the Tables.

**Figure 2a,b** Galaxy surface brightness profiles in the I band. We show the $r^{1/4}$ bulge (dashed line), exponential disk (dotted line) and combined fit (solid line) to each profile. The galaxies are numbered as in the Tables.

**Figure 3** Spectra of our sample of 18 faint (I < 22) field galaxies. Each spectrum has been smoothed to the instrumental resolution, and has the derived redshift labeled in the upper left (see Table 2). Lines used in the redshift determination are denoted by a black dot. Relative counts are given on the y axis (the spectra are not flux calibrated).

**Figure 4a,b** Velocity line profiles for H$\beta$ and [OIII]5007Å . The data for the Sb galaxy #12 ($M_B = -20.1$, z = 0.256) are shown by filled circles. The same broadened Gaussian is represented by a solid line for both the H$\beta$ and [OIII] line profiles. The instrumental profile is represented by a dotted line. After correcting for the instrumental profile in quadrature, the FWHM = 247 km s$^{-1}$.

**Figure 5** Total V–I color vs total I band magnitude. The magnitude–limited WFPC2 sample (I < 21.7) of Phillips et al. (1995b) is shown as small circles, and our sample as large open circles. Three galaxies are common to both samples. Errors on individual

measurements are $\sim 0.15^m$. Both our sample and that of Phillips et al. have a median color of 1.0. For comparison purposes, a typical local Sbc spiral has rest frame V–I color of 1.1 (Frei & Gunn 1994).

**Figure 6** Angular half light radius vs total I band magnitude. The symbols are the same as Fig. 5. Errors on individual measurements are 0.15%. Our sample has a median half light radius of $0.9''$, whereas it is $0.75''$ for the Phillips et al. (1995b) sample.

**Figure 7** Morphological Type vs total I band magnitude. The magnitude–limited WFPC2 sample of Forbes et al. (1994) is shown as small circles. This sample is 70% complete to I = 22, and essentially 100% complete to I = 21. Our current sample is shown by large open circles. There are 10 galaxies in common. Here, E = –5, E/S0 = –3, S0 = –2, S0/a = 0, Sa = 1, Sb = 3, Sc = 5, Sd = 7, Irregulars = 10, bulge–dominated = –4, intermediate = –1 and disk–dominated = 4. Barred spirals are included with their spiral subtype; double nucleus galaxies are not shown.

**Figure 8** Luminosity–disk size relation. The exponential disk scale length is plotted against rest frame absolute B magnitude of the disk. Filled symbols represent galaxies that are clearly spiral in nature, open symbols for other galaxies. The sample is also divided by intrinsic color, with circles for blue galaxies and triangles for red galaxies. Errors are estimated to be 15% in the scale length and $0.2^m$ in absolute magnitude. The magnitude error bar is shown in the lower right. Galaxies from the $\bar{z} \sim 0.3$ sample of Colless et al. (1994) are shown by small open squares. The solid and dashed lines represent the local relation assuming a central surface brightness of $21.45 \pm 0.76$ B mag arcsec$^{-2}$ (see section 4.5).

**Figure 9** Luminosity–internal velocity relation. The velocity line width sigma ($\sigma = 2\,V_{max}/2.35$) is plotted against total absolute B magnitude of the galaxy. The same symbols are used as in Fig. 8. Two disk dominated galaxies (z ∼ 0.2) from Vogt et al. (1993) are shown by small open squares. The upper solid lines represent the local relation for Sa and Sc galaxies from Rubin et al. (1985), with dashed lines giving the upper and lower bounds. The HI Tully–Fisher relation would lie between Sa and the Sc galaxy relations. The lower lines represent the local relation for HII galaxies from Telles (1995).